\documentstyle[fleqn,amssymb]{tp}

\input psfig

\newcommand\ba{\begin{array}}
\newcommand\ea{\end{array}}
\newcommand\bc{\begin{center}}
\newcommand\ec{\end{center}}
\newcommand\be{\begin{enumerate}}  
\newcommand\ee{\end{enumerate}}  
\newcommand\bi{\begin{itemize}}  
\newcommand\ei{\end{itemize}}  
\newcommand\bd{\begin{description}}  
\newcommand\ed{\end{description}}  
\newcommand\beq{\begin{equation}}  
\newcommand\eeq{\end{equation}}  
\newcommand\beqa{\begin{eqnarray*}}  
\newcommand\eeqa{\end{eqnarray*}}

\newcommand{\fig}[1]{Fig.\ \ref{#1}}
\newcommand{\etal}{{\em et al.\ }}
\newcommand\cf{{\em cf\ }}
\newcommand\eg{{\em e.g.\ }}

\newcommand\ie{{\em i.e.\ }}

\newcommand\mathC{\mkern1mu\raise2.2pt\hbox{$\scriptscriptstyle|$}
                {\mkern-7mu\rm C}}

\def\exl{\raise1pt\hbox{$\scriptstyle<$}}
\def\exr{\raise1pt\hbox{$\,\scriptstyle>$}}

\renewcommand \fig[1]{Figure~\ref{#1}}

\begin{document}

\twocolumn[
\title{CMB anisotropy power spectrum statistics}
\author{Benjamin D. Wandelt, Eric Hivon, Krzysztof M. G\'orski\\
{\it Theoretical Astrophysics Center, Juliane Maries Vej 30, DK-2100 Copenhagen}\\
}
\vspace*{16pt}   

ABSTRACT.\
Much attention has been given to the problem of estimating
cosmological parameters from the $C_l$ measured
by future experiments. Many of the
approaches which are being used
either invoke poorly controlled
approximations or are computationally
expensive. We derive exact results as well
as fast and highly accurate approximations
for mapping a theoretical model onto the
observed power spectrum coefficients and
computing their statistical properties.
These results obtain from an analytic
framework which applies for any azimuthally
symmetric sky coverage regardless of the
fraction of the sky observed by the
experiment.

\endabstract]

\markboth{Benjamin D. Wandelt \etal.}{CMB anisotropy power spectrum statistics}

\small


\section{Introduction}
There has been much recent attention to the problem of estimating the
power spectrum of anisotropies in
the cosmic microwave background, 
$C_l$, from observations (Tegmark 1997; Bond, Jaffe \& Knox 1998a,b). This is
particularly topical because present and future experiments, from
the ground, from balloons and from space, promise to
provide a wealth of cosmological
information (Knox 1995; Jungman et al.~1995;Bond, Efstathiou, \&
Tegmark 1998; Zaldarriaga, Spergel, \& Seljak 1997). 
Even if we leave the
formidable task of going from a time-ordered data set to a 
pixelised map to one side and assume the associated tasks of
controlling systematic effects and foregrounds 
will be so well controllable that they can be forgotten about in the
final maps, the $C_l$ estimation still poses several fundamental
difficulties. 
These stem mainly from the following facts:
\be
\item Balloon--borne and ground--based experiments observe only a
fraction of the sky, while the theoretical $C_l$ are global quantities
which would need observation of the full sky. Only in the full sky
limit are the $C_l$ statistically independent quantities.
\item Even satellite experiments cannot observe the full CMB sky due to
galactic obscuration.
\item Any measurement will be noisy, with a possibly anisotropic noise
pattern. 
\item The sheer size of future data sets makes the estimation problem
computationally very expensive typically scaling as the cube of the
number of pixels (Bond, Jaffe, \& Knox 1998a,b).
\item Even for Gaussian theories the $C_l$ are non-Gaussian because
they are quadratic quantities. In fact they are the
sums of the squares of correlated Gaussian random variables. The
underlying Gaussian quantities are the 
mode amplitudes $a_{lm}$ of the 
spherical harmonic expansion of the sky signal.
\ee

It has been realised long ago that the exact,
theoretically trivial way of solving the problem by maximising the
Gaussian likelihood  written in terms of the measured $a_{lm}$ is
computationally not feasible.

Therefore various approaches have been used to deal with these difficulties: 
\bd
\item [Compress the data set] in a more or less lossy
fashion and then perform a full likelihood analysis in the resulting system with
fewer degrees of freedom. 
\item [Neglect items 1 and 2 above], \ie treat the  $C_l$ as
statistically independent.
This is computationally a great simplification and hence very useful
but leads to biased results.
\item [Neglect item 5 above], \ie treat the $C_l$ as Gaussian variates
whose first 2 moments are identical with the true distribution. Thanks
to the Central Limit Theorem, this is a good approximation as long as
one can firmly establish the $l$-regime where it holds to the desired
level of accuracy. The higher order moments must decay sufficiently
that the mode and the mean of the true likelihoods are coincident to a
good approximation.
\ed

It should be noted that Oh, Spergel, \& Hinshaw (1998) have
demonstrated an  algorithm for solving the likelihood
problem efficiently by optimising their calculations explicitly 
for the MAP satellite. This achieves a computational cost
which scales with the square of the number of pixels. They also treat
for the first time
the masking of a small percentage of the CMB sky by point
sources. However, we believe that it is fair to say that their methods
 rely on the
large sky coverage of a space mission for good performance and 
are still far from computationally trivial.

The work presented in this talk presents a fresh look at the
problem. The goal is to step on to the path towards providing a
mathematical and computational tool
which satisfies both the criterion of accuracy to the exacting
standards of tomorrows CMB experiments and is computationally feasible
for a large and interesting class of cosmological theories, 
experimental setups and
observational strategies.

\begin{figure*}
\centering\mbox{\psfig{figure=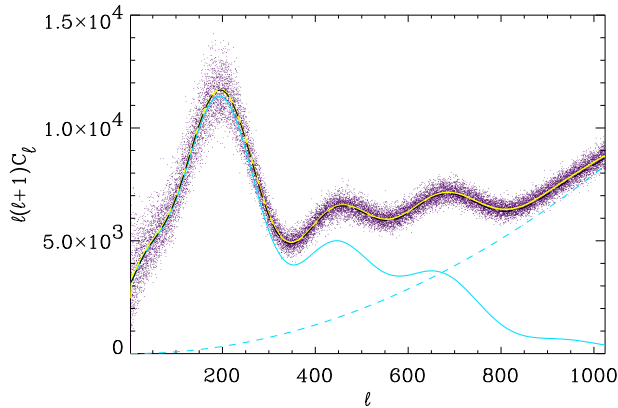,height=6cm}}
\caption[Correlation bias for an almost full sky experiment]{We show
the theoretical standard cold dark matter 
$C_l$ spectrum (solid grey line) superposed on 20 of our
Monte Carlo simulations (dots). The white noise spectrum (grey
short-dashed line) grows as
$l(l+1)C_N$ in our units. The long-dashed black line shows the sum of
signal and noise while the white solid line displays the $\langle \tilde{C}_l \rangle$ 
from 3328 Monte Carlo simulations. Apart from the large error due to
non-Gaussianity  at small $l\lesssim 30$, 
a small discrepancy can be seen for all
$l$ even for this COBE-style, almost full sky observation (\cf also
\protect\fig{moments}).}
\label{pclplot}
\end{figure*}

\section{Notation}
The full sky of CMB temperature fluctuations can
be expanded in spherical harmonics, $Y_{lm}$,
as\footnote{We assume that there is insignificant signal power in modes
with $l>l_{max}$ and use the convention that sums over $m$ run from
$-l_{max}$ to $l_{max}$ and all quantities with index ${lm}$ vanish
for $m>l$.}
\begin{equation}\label{alms}
T({\bf \gamma})=\sum_{l=0}^{l_{max}}\sum_{m}a_{lm}Y_{lm}(\gamma)
\end{equation}
where $\mathbf{\gamma}$ denotes a unit vector pointing at polar angle $\theta$ and
azimuth $\phi$.
A Gaussian cosmological theory states that the $a_{lm}$ are Gaussian
distributed with zero mean and specified variance 
$C^{theory}_l\equiv\langle\left\vert {a_{lm}}\right\vert ^2\rangle$. 
Hence, for noiseless, full sky measurements,
each measured $C_l$ independently follows a
$\chi^2$--distribution with $2l+1$ degrees of freedom and mean $C^{theory}$.

Owing to Galactic foregrounds, limited surveying time or other
constraints inherent in the experimental setup, the
temperature map that comes out of an actual measurement will be
incomplete. In addition, a given scanning strategy will produce a
noise template. We model the noise as a Gaussian field $T_N$
with zero mean which is independent from pixel to pixel and modulated
by a spatially varying rms amplitude $W_N({\mathbf{\gamma}})$. Therefore the
observed temperature 
anisotropy map is in fact
$$
\tilde{T}({\mathbf{\gamma}})=W({\mathbf{\gamma}})\left[T({\mathbf{\gamma}})+W_N({\mathbf{\gamma}})T_N({\mathbf{\gamma}})\right]
$$
where $W$ is unity in the observed region and zero elsewhere.

Expanding $\tilde{T}$ as in 
Eq.\ (\ref{alms}) produces a set of {\em correlated} Gaussian
variates $\tilde{a}_{lm}$ for the signal and $\tilde{a}_{N\;lm}$ for the
noise. These combine into
power spectrum coefficients
\begin{equation}
\tilde{C}_l=\frac1{2l+1}\sum_m
\left\vert{\tilde{a}_{lm}+\tilde{a}_{N\;lm}}\right\vert ^2
\label{pcldef}
\end{equation}
whose statistical properties differ from the ones of the $C_l$. Hence
we refer to the $\tilde{C}_l$ as {\it pseudo--}$C_l$.

\section{Results: $C_l$ Statistics}
The approach we take is to work out the complete statistical 
predictions of Gaussian
theories of structure formation for the results obtained by a given
experiment. The results we list in the following hold exactly for the case
of azimuthally symmetric sky coverage (such as rings, annuli, polar
caps, or a combination of the above, and ``full'' sky with a galactic
cut) and a white noise distribution which is allowed to be anisotropic but
must have the same axis of symmetry as the observed sky
region. However, we
show by comparison with extensive Monte Carlo simulations that our
results are miraculously accurate even for strongly misaligned,
symmetry violating noise
patterns, far into the noise dominated regime (so in this case read
``approximate'' in place of ``exact''). 

We emphasize that while our Monte-Carlo simulations are computed for
a $\pm 20^\circ$ galactic cut, our approach is {\em not}
dependent on large sky
coverage. We chose to compute this  case as a demonstration
of the accuracy of our formalism, its remarkable robustness with respect
to a tilted noise template and its computational feasibility for
 maps with several million pixels. For such maps this is,
to our knowledge,  the only rigourous
method which allows the discussion of such parameter biases and the
computation of
$C_l$ statistics to the levels of accuracy which are quoted as the
baseline for \eg the Planck mission (Bersanelli et al. 1996). 

Here are the results one by one (Wandelt, Hivon, \& G{\'o}rski 1998a,b):
\be
\item An exact generalisation of the $C_l$ sample variance formula
(Knox 1995) to partial sky coverage
and anisotropic noise.

\begin{figure*}[t]
\centering\mbox{\psfig{figure=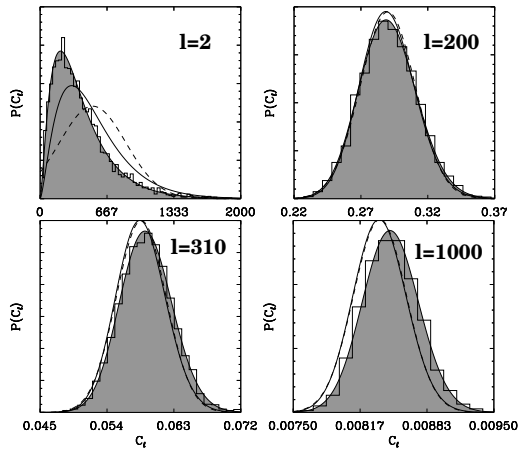,height=6cm}}
\caption{The $\tilde{C}_l$ distributions. We compute them 
for the  standard cold dark matter model (shaded), 
the $\chi^2$ (solid lines) and the Gaussian (dashed) approximations
compared to Monte 
Carlo simulations (histograms) for
l=2,200,310,1000.}
\label{pcldists}
\end{figure*}

\item An exact analytical closed form 
solution for the non-Gaussian marginal probability distributions of
the $\tilde{C}_l$. Examples of these distributions are shown in \fig{pcldists}.

\begin{figure*}[t]
\centering\mbox{\psfig{figure=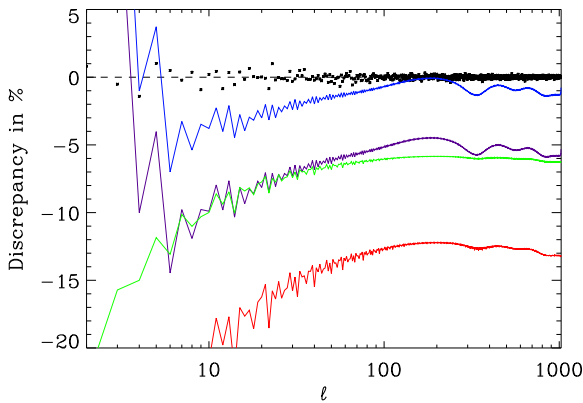,height=6cm}}
\caption[]{Moment discrepancies between the WHG and $\chi^2$
distributions. We show the discrepancy in percent between the means,
standard deviations, skewness and kurtosis for the $\tilde{C}_l$ and 
$\chi^2$ distributions. The stars are the $\langle \tilde{C}_l \rangle$ computed
from 3328 Monte Carlo simulations, showing excellent agreement only
limited by Monte Carlo noise. Notice the oscillatory features in the discrepancy
which indicate that it is dependent on the underlying theory, \eg through
the position and amplitude of the peaks. The difference in the means
for $l \gtrsim 30$
is dominated by the correlation bias and does not decay for large $l$. }
\label{moments}
\end{figure*}

\item Exact expressions for cumulants and joint cumulants (and hence
moments) of all orders
for the $\tilde{C}_l$. In \fig{moments} we compare the true 
means, standard deviation, skewness
and kurtosis  to the ones for the $\chi^2$ distribution for all
$l$. These can form the basis of a debiasing scheme, as suggested in 
Wandelt Hivon \& G{\'o}rski (1998a).

\item A new, standardised, $\chi^2$ distributed 
statistic $y_l$ which measures the goodness of fit between a set of
observed $\tilde{C}_l$ and a given theory\footnote{We thank G. Efstathiou for
suggesting computation of this quantity to us.}.

\begin{figure}
\centering\mbox{\psfig{figure=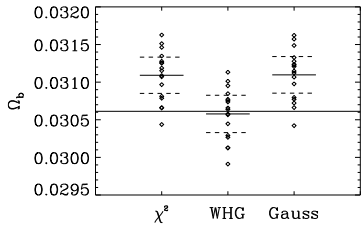,height=3.75cm}}
\caption[]{Correlation bias in cosmological parameter estimation. We
estimate the baryon fraction in the universe from 17 realisations of
$\tilde{C}_l$, comparing the Gaussian and $\chi^2$ approximation to
our likelihood (WHG). The true value is marked with the long solid
horizontal line. The mean estimates are marked
with short horizontal lines. Because we enforce the correct marginal
distributions our result is guaranteed to be unbiased while the
Gaussian and $\chi^2$ approximations neglect $l$ to $l$ correlations
and hence produce a systematic bias, here detected at more than 3
sigma (dashed horizontal lines). }
\label{obbias}
\end{figure}

\item The discovery and removal of a {\em correlation}  bias
in cosmological parameter estimation 
which results  from neglecting the correlations induced
by partial sky coverage and anisotropic noise. This is an effect which
is distinct from and
in addition to the bias which comes from neglecting the
non-Gaussianity of the $\tilde{C}_l$. This non-Gaussianity bias has
been termed ``cosmic'' bias elsewhere 
(Bond, Jaffe, \& Knox 1998b). The  correlation bias occurs at all 
$l$  even those where the 
Central Limit Theorem ensures Gaussianity in the case of almost full
sky coverage. 
\footnote{Of course our calculations
also account for the non-Gaussianity of the $\tilde{C}_l$ distributions and
are therefore free from ``cosmic'' bias.}. To show the effect on the
estimation of cosmological parameters we form an approximate
likelihood by multiplying the marginal $\tilde{C}_l$-distributions and
use this to estimate
$\Omega_b$, the baryon fraction for the standard cold dark matter
scenario. The results are shown in \fig{obbias}. While this is quite a
small effect for the almost full sky case considered in
Wandelt, Hivon \& G{\'o}rski (1998a) we conjecture this to become more important for
medium to small sky coverage.
\ee

The computation of any of these quantities, except for the joint
moments, scales strictly as
$N_{pix}^{3\over2}$, where $N_{pix}$ is the number of pixels, with a
small prefactor. This means all computation take on the order of 1
minute for an $l$-range of 0 to 1024. In the case of the joint moment
$\left\langle\Delta\tilde{C}_l\Delta\tilde{C}_{l'}\right\rangle$ the
computation scales as $N_{pix}^2$ and takes on the order of a few
minutes.

Work is in progress  to extend these methods to a viable
tool for the joint estimation of the Big Bang $C_l$
and cosmological parameters in Gaussian theories from noisy,
cut-sky data (Wandelt, Hivon \& G{\'o}rski 1998b).

\section*{Acknowledgments}

We would like to thank A.\ J.\ Banday for stimulating discussions. This
work was funded  by the Dansk Grundforskningsfond through its
funding for TAC.



\end{document}